\documentstyle[twocolumn,prl,aps,psfig]{revtex}
\begin{document}
\twocolumn[\hsize\textwidth\columnwidth\hsize\csname
@twocolumnfalse\endcsname

\title{Two-Leg Ladders and Carbon Nanotubes: Exact Properties at Finite Doping}
\author{R. Konik$^1$, F. Lesage$^2$, 
A.W.W. Ludwig$^1$, H. Saleur$^3$.}
\address{
$^1$Department of Physics, UCSB, Santa Barbara, CA 93106}
\address{
$^2$Centre de recherches math\'{e}matiques, Universit\'{e} de Montr\'{e}al,
C.P. 6128, Succ. centre-ville, Montr\'{e}al H3C 3J7, Canada}
\address{
$^3$Department of Physics, University of Southern California,
Los-Angeles, CA 90089-0484}
\date{\today}
\maketitle
\begin{abstract}
Recently Lin, Balents, and Fisher have demonstrated that two-leg
Hubbard ladders and armchair carbon nanotubes renormalize
onto the integrable SO(8) Gross-Neveu model. 
We exploit this integrability to examine these systems in their doped
phase.  Using thermodynamic Bethe ansatz, we 
compute exactly both the spin and single particle gaps and
the Luttinger parameter describing low energy excitations.
We show both the spin gap and particle gap do not vanish at
finite doping, while the Luttinger parameter remains close to its
free fermionic value of 1.  A similar set of conclusions is
drawn for the undoped systems' behaviour in a finite magnetic field.
We also comment on the existence
in these systems of the $\pi$-resonance, a hallmark of Zhang's
SO(5) theory of high $T_c$ superconductivity.

\end{abstract}
\pacs{PACS numbers: ????}

]

\newcommand{\del}{\partial}
\newcommand{\nn}{\nonumber}
\newcommand{\rtc}{\tilde{\rho}_c}
\newcommand{\th}{\theta}
\newcommand{\gcc}{\Gamma_{cc}}

Carbon nanotubes are among a new class of quasi one-dimensional materials
exciting tremendous interest in both the material science and physics
communities\cite{nanorefs}\cite{kane}\cite{balents}.
Nanotubes are built by wrapping graphite sheets into
tubes of nanometer scale.  They take a multitude of forms, ranging from
crystalline ropes of nanotubes arranged in triangular lattices to single
multi-walled nanotubes consisting of concentric cylinders to the simplest
construction, a single-walled, isolated nanotube.  The latter promises
to provide cleaner and longer realizations of quantum wires and as such
promises to be the most relevant in new technologies\cite{kane}.

Two-leg Hubbard ladders form another prominent class of quasi 
one-dimensional systems.  Interest in these systems arises partly
from the novel spin physics one expects in coupled chains\cite{dagotto},
and partly in that they provide an example of a Mott-insulator, 
a class of compounds that  
includes undoped high $T_c$ cuprates.  In contrast to the
latter, here long-range antiferromagnetic order cannot occur 
and there is a finite gap 
to spin excitations.  As such they 
represent a spin liquid.
Their basic doped behaviour has been
examined both through weak coupling analytical treatments 
\cite{analy}\cite{fisher}
and through Monte Carlo and density matrix renormalization group (DMRG)
techniques\cite{num}: the spin gap decreases in magnitude upon doping
and the system
exhibits quasi long range d-wave superconducting pairing correlations.
We analyze this behaviour exactly in the following.

Remarkably a certain class of the single walled nanotubes behave as 
two-leg Hubbard ladders\cite{balents}.
Single walled tubes are classified by a superlattice
translation vector, denoted by (N,M), specifying how the graphite sheet is 
to be rolled up in forming the tube.  ``Armchair'' tubes, described by
a superlattice vector (N,N), have been shown to be identical to a two-leg 
ladder with an effective Hubbard interaction of $U/N$.  Our
findings for doped ladders thus apply equally to doped armchair nanotubes.
Due to their relative cleanness, 
it may well be in such systems that our results are most readily
observed.  Electronic properties of the tubes
can be easily measured by attaching
metallic leads or by tunneling through an STM tip.  The feasibility
of the latter has been recently demonstrated by J. 
Wild\"oer et al.\cite{nanorefs}.  Exact results for the tunneling
$I(V)$-curve into an infinitely long tube have recently been obtained by
two of us\cite{rmk}.

Even more remarkable is the finding by Lin, Balents, and Fisher \cite{fisher}
that the two-leg Hubbard ladder, and thus the armchair nanotubes, 
renormalize
in weak coupling onto an SO(8) Gross-Neveu model.  This
equivalence is established by considering {\it generic} interactions.
Under a one-loop renormalization group flow, the interacting Hamiltonian flows
onto an SO(8) Gross-Neveu model,

\begin{eqnarray}\label{e1}
H &=& \int dx \bigg(\sum^4_{a=1} \big(\Psi^\dagger_{aL}i\partial_x\Psi_{aL} 
- \Psi^\dagger_{aR}i\partial_x\Psi_{aR} \\
\nn && \hskip .4in + g (\Psi^\dagger_{aL}\Psi_{aR}
- \Psi^\dagger_{aR} \Psi_{aL})^2 \big)\bigg).
\end{eqnarray}
a theory of four interacting Dirac fermions,
$\left( \psi_{Ra} , \psi_{La} \right)$.  

SO(8) is an immense symmetry, usually found only in the domain of 
grand unified
theories and supersymmetric strings\cite{witten}.  
In condensed matter systems we are 
usually limited to considering systems
with charge conservation (U(1) symmetry) and 
perhaps rotational invariance (SU(2) symmetry).  
But we are doubly blessed.  The SO(8) symmetry resides in an
integrable model.  Often integrable models are considered to
be of little use in describing experiments in that they are
highly tuned and thus are not expected to
describe generic behavior in
experimental systems.  However here we avoid this difficulty:
the system is attracted to the integrable SO(8) Gross-Neveu
model under a renormalization group flow.  A similar situation
was found in the tunneling between $\nu\!=\!1/3$ quantum Hall
edge states\cite{edges}.

Integrability provides us with a tremendous amount of information.
The key to organizing this information is the exact knowledge of the
basis of eigenstates of the fully interacting system.  There is a precise
notion of a ``particle'' in an integrable system akin to fermi liquid
theory.  However unlike fermi liquid theory the particle is perfectly
stable.  There is no particle decay or production; all scattering processes
are encoded in two particle scattering matrices, $S$.

Particles in the SO(8) Gross-Neveu model are organized into three octets
and a set of 29 bosonic particles.
One octet is of fundamental (Majorana)
fermions
and is directly related to the particle/hole excitations associated with
the four Dirac fermions in eqn.(\ref{e1}).  The remaining two octets are of
fermionic ``kink'' particles.  The 29 bosons consist of a rank two anti-symmetric
tensor with a single scalar boson, together forming a single representation
of the SO(8) Yangian.  The masses of these particle are generated
dynamically through the Gross-Neveu interactions (the quartic term in
eqn.(\ref{e1})).  In terms of the ladder parameters, the mass scale, $m$, is
governed by, $m \sim te^{-t/U}$, where $t$ is the ladder bandwidth and $U$
is the typical coupling strength.  The fermions all have mass, $m$, whereas
the bosonic particles have mass, $\sqrt{3}m$.

Knowledge of these masses together with knowledge of their quantum numbers
was used to make a variety of predictions in \cite{fisher} with respect
to the correlation functions at low temperature and zero doping.  
Largely
qualitative, these results were extended in \cite{rmk}, where matrix elements
were computed exactly in order to obtain exact forms for the correlators.
Here we consider the doped
system at zero temperature.  Though more difficult theoretically, it is more
interesting experimentally.  Beyond a critical value of the chemical potential,
the gap to charge-two excitations goes
to zero and the system becomes conducting.

To model the doping, we add a chemical potential term, $H_Q = -\mu Q$, to
the Gross-Neveu Hamiltonian where $Q$ is the total electric charge.  As $Q$
is conserved by the original Hamiltonian, this addition
does not spoil integrability.  Rather it only lifts the degeneracy
of particles with differing charges.  To determine the charge of the
octet of fundamental (Majorana) fermions, we organize them into four
Dirac fermions (only in doing this do they have well defined charges).
Of the four, one carries electric charge 2, the so-called cooperon,  
and one carries spin $S_z=1$, the magnon.
All four are associated
with two-electron excitations.  While the fundamental fermions
carry but a single quantum number, the two octets of kink particles each
carry $\pm 1/2$ of each of the four above quantum numbers.  In 
particular, kinks
carry charge, $Q=\pm 1$, and spin, $S_z=\pm 1/2$, 
and so a subset of them are identified
with the physical electronic excitations on the ladder\cite{fisher}.  
The 29 bosonic
particles can be thought of as bound states of the fundamental fermions,
and so carry two quantum numbers, for example, $Q=2$ and $S_z=1$.

As the chemical potential is turned on, a particle with charge $Q$ lowers
its energy to $m-\mu Q$.  As the energy
of a particular particle becomes negative, the ground state of the system
changes.  Once empty, it now begins to fill with particles,
much as if the fermi energy was moved above the bottom of the conduction
band in a semi-conductor.  The first particle that will enter the ground
state is the one with the highest charge to mass ratio.  This is the
cooperon.  The ground state is pictured in Fig. 1 for $2\mu > m$,
where it consists solely of cooperons. 

\begin{figure}[tbh]
\centerline{\psfig{figure=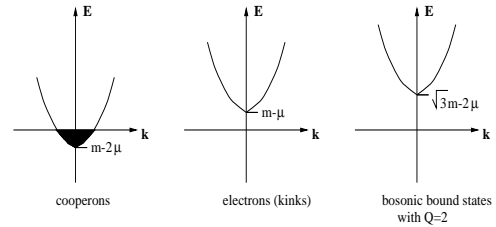,height=1.15in,width=2.5in,angle=-90}}
\caption{Schematic of the doped ground state.}
\end{figure}

\noindent Seemingly as $\mu$ is increased further, the kinks and bosonic
bound states will begin to fill into the ground state.  However the increase
in chemical potential is counteracted by the repulsion felt by the kinks and
bound states to the cooperons.  It turns out that this repulsive energy
is greater than the chemical potential {\it for arbitrary} $\mu$, and
the ground state remains solely filled with cooperons, as follows
from the general results of \cite{hollowood}.
We emphasize that this is an exact result.

We now proceed to characterize quantitatively the ground state
through thermodynamic Bethe ansatz (TBA).  At the root of the ansatz 
lies the insistence that the cooperon
wave function 
be consistent with the scattering amplitudes, $S_{cc}$ (given in
\cite{weisz}), of two
cooperons:  
\begin{equation}
\!S_{cc}\!(\th ) \!=\! {\Gamma (1\!+\!{i\th\over 2\pi}) 
\Gamma (1/2\!-\!{i\th\over 2\pi})
\Gamma(5/6\!-\!{i\th\over 2\pi})\Gamma (1/3\!+\!{i\th\over 2\pi}) \over
\Gamma (1\!-\!{i\th\over 2\pi}) \Gamma (1/2\!+\!{i\th\over 2\pi})
\Gamma(5/6\!+\!{i\th\over 2\pi})\Gamma (1/3\!-\!{i\th\over 2\pi})}.
\end{equation}
In this way we can derive a density of occupied states for
the cooperons, $\rtc$, and a corresponding energy, $\epsilon_c$, of the
cooperons.  We parameterize
the non-interacting energy and momentum of a particle of mass, $m$,
by $E=m\cosh (\th )$, $p=m\sinh (\th )$.  
With this parameterization, the
use of periodic boundary conditions results in an equation which
determines the density of occupied states per unit length for $|\th|<B$:
\begin{equation}\label{dos}
\rtc\!(\th )\!=\!{m\over 2\pi}\cosh (\th ) +\! \int^B_{-B} d\th \rtc (\th ')
\gcc (\th - \th ').
\end{equation}
Here $\gcc =\frac{1}{2\pi i}\partial_\theta \ln S_{cc}$
is a measure of the strength of the cooperon-cooperon interaction; if
$\gcc = 0$ we recover the density of occupied states for free fermions.
$\pm B$ marks out the fermi energy.  For $|\th| > B$, $\rtc (\th ) =0$.
One determines the {\it interacting} cooperon energy, $\epsilon_c (\th )$,
by associating it with the occupied density of states at $T\neq 0$
via $\rtc \equiv 1/(1+\exp(\epsilon_c/T))$.  Demanding that the
free energy be minimized and then taking $T\rightarrow 0$, 
one obtains $\epsilon_c$ in terms
of the cooperon-cooperon S-matrix:
\begin{equation}
\epsilon_c (\th ) - \int^B_{-B}d\th' \gcc (\th-\th')\epsilon_c (\th ')
= m\cosh (\th ) -2\mu.
\end{equation}
Solving the constraint, $\epsilon_c (B ) = 0$, determines $B$.

Having described the ground state, we now look at excitations over the ground
state, that is the gaps.  We have already seen 
that the gap to a  charge-two
excitation is zero; we can add a cooperon at an infinitesmial energy
above the fermi energy.  More interesting are the gaps to spin 1 and single
particle excitations.  As emphasized above,
these gaps never vanish at finite doping, a rigourous result.

We first consider the spin gap, $\Delta_s$.  
To compute this gap we want to identify
the Q=0, $S_z = 1$ excitations.  There are two possibilities:
1) adding a $Q=0$, $S_z=1$ magnon; and 2)
removing a cooperon from the sea and adding a 
bosonic bound-state with $Q=2$ and $S_z=1$.
For $2\mu < m$, the first excitation is the only one possible; the 
sea is empty.  The spin gap is then $\Delta_s = m$.
As we cross the threshold, $2\mu = m$, the Fermi sea begins to fill.
At the threshold, the second process has a gap of only $m(\sqrt{3}-1)$, the 
energy of the cooperon hole added to the energy of
the bound state \cite{fisher}.  It is thus
lower than the first process, and 
the spin gap is determined by this excitation.  
We can push beyond threshold and compute the gap for
all values of the doping: previous studies \cite{weisz}\cite{weig}
have shown how to obtain the exact ground state of the 
theory at finite doping.  
The lowest $S_z=1$, $Q=0$ excitation is given by 
removing a cooperon at the fermi surface with rapidity, $\th_h = B$, 
and adding a bound
state with rapidity $\th_b=0$.   With these changes the
particle density is modified to 
$\tilde{\rho}_c(\theta|\theta^h,\theta^b)$.  
$\tilde{\rho}_c(\theta|\theta^h,\theta^b)$
is determined by an equation similar to (\ref{dos}),
\begin{eqnarray}
\label{shlin}
\tilde{\rho}_c(\theta|\theta_h,\theta_b)-\int_{-B}^B d\theta' 
\Gamma_{cc}(\theta-\theta')
\tilde{\rho}_c(\theta'|\theta_h,\theta_b)=\nonumber \\ 
\Gamma_{cb}(\theta-\theta_b)-\Gamma_{cc}(\theta-\theta_h).
\end{eqnarray}
Here we have used $\Gamma_{cb}=\frac{1}{2\pi i}\del_\th\ln S_{cb}$, where
\begin{equation}
S_{cb} (\th ) = S_{cc}(\th - {i\pi\over 6})S_{cc}(\th + {i\pi\over 6})
{\th + {i\pi\over 6} \over \th - {i\pi\over 6}}.
\end{equation}
$S_{cb}$ is the cooperon-bound state
S-matrix\cite{rmk}\cite{karowski}.

Having the new density, the energy of the excited
state is then straightforward to compute: it is determined by
the difference between the energy of the excited system and the
energy of the original ground state.  
Given that the energy of the cooperon 
sea is $L \int d\th \rtc (\th )(\cosh(\th ) - 2\mu /m)$,
the spin gap equals
\begin{equation}
{\Delta_s\over m} = \sqrt{3} -\cosh B+\int_{-B}^B
d\theta ~\delta\tilde{\rho}_c(\th )(\cosh\theta - 2{\mu\over m}),
\end{equation}
where $\delta\tilde{\rho}_c(\th ) = \rtc(\theta|\theta_h,\theta_b)
- \tilde{\rho}_c(\th)$.

The single particle gap, defined as the energy to add an electron 
with Q=1 (i.e. a kink)
to the system, follows from similar considerations.
For $2\mu < m$, the gap is that of adding a kink in the absence of a
fermi sea, $\Delta_f = m-\mu$.
Crossing the threshold, $2\mu > m$, 
the sea begins to fill in and the interactions between the kink and
the cooperons must be taken into account.
The scattering between
the cooperons and kinks, $S_{ck}$ is given by
\begin{equation}
S_{ck} (\th) = {\th-{i2\pi\over 3} \over \th + {i2\pi\over 3}}
{s(\th ) + {i\sqrt{3}\over 2} \over s (\th ) - {i\sqrt{3}\over 2}}
{\Gamma({5\over 6}+{i\th\over 2\pi})\Gamma({4\over 3}-{i\th\over 2\pi})
\over
\Gamma({5\over 6}-{i\th\over 2\pi})\Gamma({4\over 3}+{i\th\over 2\pi})},
\end{equation}
with $s(\th)=\sinh(\th)$.  The same 
procedure as before leads to the single particle gap, $\Delta_f$:
\begin{equation}
{\Delta_f\over m} = 1 - {\mu\over m} + \int_{-B}^B
d\theta ~\delta\tilde\rho_c(\th ) (\cosh (\th ) - 2{\mu \over m}).
\end{equation}
Here $\delta\tilde\rho_c$
is given by removing $\gcc$ and
replacing $\Gamma_{cb}$ by $\Gamma_{ck} = \del_\th \ln S_{ck}/2\pi i$,
both on the r.h.s. of eqn. (5).

Both of these gaps are shown in Fig. 2.  The gaps have been plotted
vs. $D = \int d\th \rtc (\th)$, 
the density of cooperons (that is, the carrier density).
We observe that there
is a uniform decrease with increasing doping, as expected
for this system.  For large $D/m$, the gaps approach zero asymptotically
without crossing, as can be established analytically.

\vskip -.5in
\begin{figure}[tbh]
\centerline{\psfig{figure=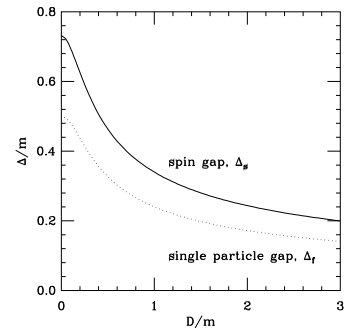,height=2.3in,width=2.3in}}
\caption{Plot of the spin and single particle gap.}
\end{figure}

\noindent
Although the gap to charge two excitations is zero, these excitations still
encode interesting physics.
The massless charge excitations occur
near
the fermi rapidity, $\theta=\pm B$.  These excitations can be
described by two chiral bosons, $\phi_{L/R}$, (one for each fermi
point).  That the cooperon sea is interacting is reflected 
in the Luttinger parameter, $K$, that governs these massless bosonic
excitations.  The propagator of $\phi_{L/R}$ is given in terms
of $K$.  Focusing on $\phi_L$ we have,
$\langle \phi_L(z)\phi_L(w)\rangle
=-K\ln(z-w)$.  The operator creating a Cooper pair is thus
$O_c=e^{i\phi_L/K}$, and so is governed by the propagator
\begin{equation}
\langle O_c(z)O_c(w)\rangle= (z-w)^{-{1 \over K}}\label{coopcor},
\end{equation}
where $z=v_F\tau + i x$.
$K$ thus describes the long distance asymptotics of this correlator.

We can easily determine K in terms of the renormalized charge, 
$Q_R = -\del_\mu \epsilon_c |_{\th=B}/2\equiv Z$, of
the elementary excitations near the fermi surface.
The basic idea, discussed
at length in \cite{ls}, is that the low energy excitations
near the Fermi rapidity are just free fermions - that is, the dressed
S matrix
at low energy is just $-1$. This means that the only non vanishing
matrix elements of the current operator, $j_L= \partial_z\phi_L$, 
in this basis involve
particle-hole pairs, which, parameterizing the energy of the
{\it massless} excitations by $E(\th ) = e^\theta$, are given by,
\begin{equation}\label{ff}
\langle \th_{\rm hole},\th_{\rm particle}| j_L(0)|0\rangle=c
e^{\th_{\rm hole}/2}e^{\th_{\rm particle}/2}.
\end{equation}
The constant, $c$, is related to the interaction
parameter, $K$, as we expect $\langle j_L (z) j_L (0)\rangle = -K/z^2$.
Using the matrix element to compute this correlator, we find
$K=-c^2/4\pi^2$.
On the other hand, $c$ is related via $c=i2\pi Q_R$ 
to the renormalized charge,
since integrating (\ref{ff}) over all $x$ gives the value of $Q_R$ 
on the one particle states.  One thus finds $K=Q_R^2=Z^2$.
$K$, using eqn. (4), is then readily computed with the results plotted in
Figure 3.  With $K<1$, the effective interactions are repulsive 
but because the cooperon-cooperon interaction 
is weak, the Luttinger parameter does not vary significantly
from its free fermionic value of $K=1$.  Interestingly, the plot
is marked by a dip at about $D/m \sim .4$ before beginning an approach
(albeit slowly) to an asymptotic value of $K = 1$.
\vskip -.5in
\begin{figure}[tbh]
\centerline{\psfig{figure=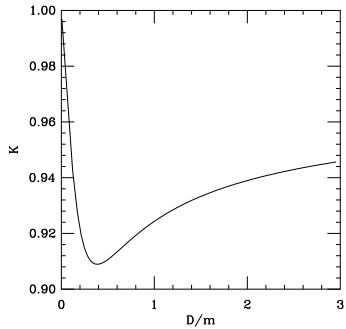,height=2.3in,width=2.3in}}
\caption{Plot of the Luttinger parameter.}
\end{figure}
We point out that all of the above analysis applies equally to the
system in a magnetic field, $H$, at zero doping.
Rather than coupling to the
cooperon, the magnetic field couples to the magnon.  However as the
magnon and cooperon are related by an SO(8) rotation, all of the
above equations hold under relabeling.  In this scenario, the
ground state consists of interacting magnons with zero spin
gap.  $\Delta_s$ becomes the {\it charge gap} while $\Delta_f$
remains the gap to single particle excitations (with $S_z = 1$).  
Here $K$ describes
the low energy spin 1 excitations about the fermi surface.

To conclude, we would like to make a few comments about the  
equivalent of the ``$\pi$ resonance'' discussed in Zhang\cite{Zhang}.
Understanding singularities quantitatively is beyond the scope of  
this note;
however, certain simple threshold effects can easily be analyzed. In  
particular, let us consider the spin spectral function.  This can be 
computed, at least in principle, by inserting a complete set of  
states between the two spin operators, corresponding to various  
physical processes. 
We consider in particular spin excitations with energy $2\mu$.  Such
excitations can be obtained by applying Zhang's $\pi$-operator to
low energy $Q=2$ excitations.  The $Q=2$ excitations are formed by adding 
a single cooperon to the ground state directly above the fermi sea 
together with cooperon particle-hole pairs.  As there are an infinite number 
of such excitations, we expect the spin-spin 
correlator to exhibit, at momentum 
($\pi$,$\pi$), a singularity 
of the type $|\omega-2\mu|^x\Theta(\omega-2\mu)$, akin to that
found in \cite{fisher}.  However unlike \cite{fisher} we expect this
singularity to be robust:
with integrability,  massive excitations well above the  
continuum can still
be stable because they are protected by higher order conserved  
quantities preventing their decay, although clearly, non-integrable
perturbations will give these excitations a small finite lifetime.

The authors would like to acknowledge discussions with H.H.Lin, 
L. Balents, M.P.A. Fisher, and D. Duffy. This work has been supported by
NSERC and the National Science Foundation through the Waterman Award under
grant number DMR-9528578 (R.K.),
by the A.P. Sloan Foundation (A.W.W.L.), the Packard Foundation, the NYI
Program, and the DOE (H.S.).

\end{document}